\documentclass[final]{aipproc}
\newcommand\selectedlayoutstyle{8x11single}
\layoutstyle\selectedlayoutstyle
\SetInternalRegister\hbadness{8000} 
\usepackage{graphicx}   
\newdimen\captwidth   
\newdimen\figwidth   
\def\eg{{\it e.g. }}
\def\etal{{et al. }}  

\def\micron{\hbox{$\,\mu {\rm m}\,$}}
\def\milli{\hbox{$\, {\rm mm}\,$}}

\begin{document}

\title 
      [The Diabolo Photometer]
      {The Diabolo photometer and the future of ground-based
        millimetric bolometer devices}


\author{Désert, F.--X. }{
  address={Laboratoire d'Astrophysique de 
l'Observatoire de Grenoble, 414 rue de la Piscine, BP53, 
F--38041 Grenoble Cedex 9, France}
}

\iftrue
\author{Benoît, A., Camus, Ph.}{
  address={Centre de Recherche sur les Tr\`es Basses Temp\'eratures,
25 Avenue des Martyrs BP166, F--38042 Grenoble Cedex 9, France}
}
\author{Giard, M., Pointecouteau, E.}{ address={Centre d'\'Etude
    Spatiale des Rayonnements, 9 avenue du Colonel Roche, BP 4346,
    F--31029 Toulouse Cedex France} } \author{Aghanim, N., Bernard,
  J.--P., Coron, N., Lamarre, J.--M., Marty, Ph.}{ address={Institut
    d'Astrophysique Spatiale, B\^at. 121, Universit\'e Paris XI,
    F--91405 Orsay Cedex, France} } \author{Delabrouille, J.}{
  address={Physique Corpusculaire et Cosmologie, College de France, 11
    pl. Marcelin Berthelot, F-75231 Paris Cedex 5} }
\author{Soglasnova, V.}{ address={Space Research Institute, Astrospace
    Center, Academy of Science of Russia, Profsoyuznaja St. 84/32
    117810 Moscow, Russia} } 
\fi \copyrightyear {2001}

\begin{abstract}
  The millimetric atmospheric windows at 1 and 2~mm are interesting
  targets for cosmological studies. Two broad areas appear leading
  this field: 1) the search for high redshift star-forming galaxies
  and 2) the measurement of Sunyaev--Zel'dovich (SZ) effect in
  clusters of galaxies at all redshifts. The Diabolo photometer is a
  dual-channel photometer working at 1.2 and 2.1\milli and dedicated
  to high angular resolution measurements of the Sunyaev--Zel'dovich
  effect towards distant clusters. It uses 2 by 3 bolometers cooled
  down to 0.1~K with a compact open dilution cryostat. The high
  resolution is provided by the IRAM 30m telescope. The result of
  several Winter campaigns are reported here, including the first
  millimetric map of the SZ effect that was obtained by Pointecouteau
  \etal (2001)~\cite{P01} on RXJ1347-1145, the non-detection of a millimetric
  counterpart to the radio decrement towards PC1643+4631 and 2~mm number
  count upper limits. We discuss limitations in ground-based
  single-dish millimetre observations, namely sky noise and the number
  of detectors. We advocate the use of fully sampled arrays of (100 to
  1000) bolometers as a big step forward in the millimetre continuum
  science. Efforts in France are briefly mentionned.
\end{abstract}

\date{\today}

\maketitle

\section{Introduction}
The atmospheric windows at 1 and 2~mm wavelengths constitute a large
opening for ground--based cosmological studies. Continuum observations
on large single--dish telescopes (IRAM 30~m, JCMT, CSO, SEST, \ldots)
have already provided outstanding results in that respect. Whereas the
search for high redshift galaxies has proved very successful in the
near past mostly at 1.2 and 0.8~mm (this conference), we would like
here to also emphasize the usufulness of millimetre SZ measurements in
the 2.1~mm window by showing the results that have been achieved with
the Diabolo instrument\footnote{More details on the experiment can be
  found at
  http://www-laog.obs.ujf-grenoble.fr/$\tilde{\,}$desert/diabolo/diabolo.html}. The
IRAM 30~m millimeter telescope at Pico Veleta (Spain) provides the
highest angular resolution on SZ effect with the combination of the
size of the telescope and the operating wavelength, namely about
20~arcsecond at 2~mm. This can be very important for the study of high
redshift clusters of galaxies which may not be fully virialized.

One can note that both (sub)millimetric flux of galaxies and the
SZ effect brightness (although not for the same reason) share the property of
being rather insensitive to their redshift. Hence, number counts can
be much more sensitive to the luminosity function than to distance
effect. The high redshift population of objects can stick out more
easily than in other wavelength domains. In particular, when confusion
is close, this can be a very important positive leverage to extract
the early population from the low redshift crowd.

\section{The Diabolo instrument}

\begin{figure}[htbp]
  \includegraphics[angle=0,height=.3\textheight]
    {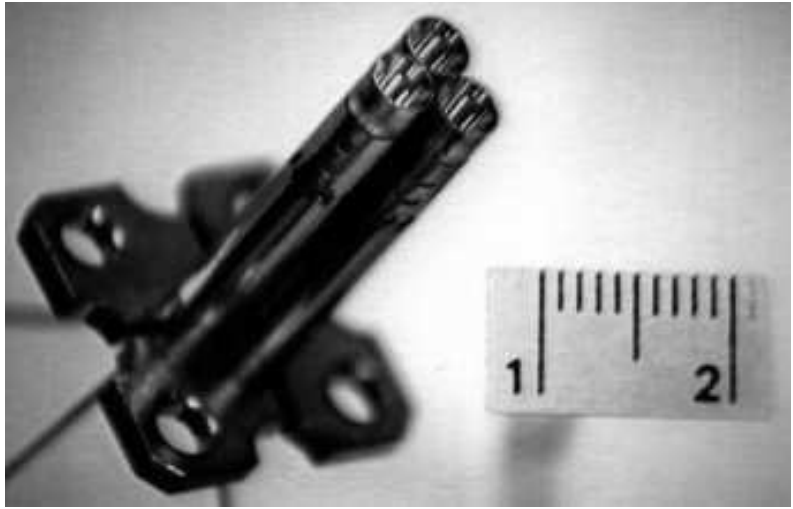}
      \caption{One of the 2 arrays of 3 bolometers used in
        Diabolo. The Winston cones are arranged in a close-packed
        triangular configuration. }
         \label{fig:3bolos}
   \end{figure}
\begin{figure}[htbp]
  \includegraphics[angle=0,height=.3\textheight]
    {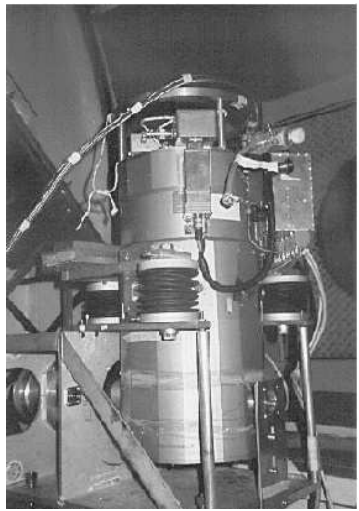}
      \caption{Diabolo cryostat in the Nasmisth cabin of the 30m. One
        can see shock absorbers (black springs) around the cryostat to
        damp vibrations coming from cryocoolers in the same cabin. The
        electronics box is on the upper right.}
         \label{fig:dbo30m}
   \end{figure}
   
   Diabolo is
   a dual-channel photometer with 0.1~K bolometers cooled by a
   space-compatible dilution fridge of the same type as what will be
   flown on Planck-HFI (Lamarre \etal, this conference). It is
   described in length by Benoît \etal (2001)~\cite{Benoit}. The AC square bias
   electronics to read the bolometers is described by Gaertner \etal
   (1997)~\cite{Gaertner}.

It now contains two small arrays of 3 bolometers each of which has a
Winston cone at its entrance aperture (Fig.~\ref{fig:3bolos}). With a
beam splitter, both arrays simultaneously measure the sky brightness
resp. at 1.2~mm and at 2.1~mm. This is essential to spectrally
separate sky noise from the SZ effect (see below). The FWHM of the
beam is 22~arcseconds when the photometer is installed on the IRAM~30m
telescope at Pico Veleta (Spain). Fig.~\ref{fig:dbo30m} shows the
cryostat at the Nasmith focus.

In 1995 and 1996, we performed ON-OFF (target) along with ON-OFF
(blank-sky) on selected clusters of galaxy (Désert \etal
1998~\cite{Desert}) to make first detections and to check for
systematics. Since then, we have done small raster maps where the
telescope is held fixed in local coordinates and the Earth rotation
makes a drift subscan at constant declination. A map is made by
repeating those subscans at different declination.

The sensitivity is below $1\times 10^{-4}$ (about 1~mJy/beam) for the
comptonisation parameter $y$ at $1\sigma$ in one hour of integration
and after sky noise is subtracted (see below).

\section{Scientific results}

\subsection{The SZ effect}

\begin{figure}[htbp]
  \includegraphics[angle=90,height=.5\textheight]
    {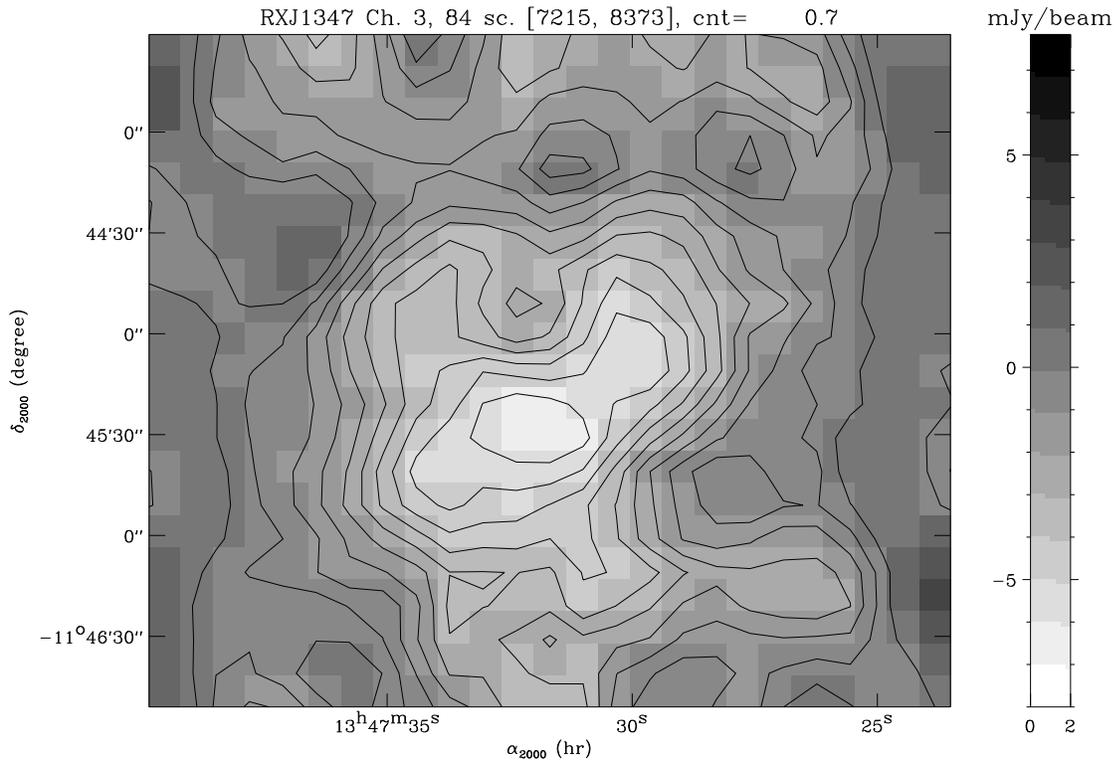}
      \caption{2.1~mm map obtained with the Diabolo instrument in
        1999 of the cluster of galaxies RXJ1347-1145. It corresponds
        to the coaddition of 84 independant rasters. The grey scale is
        from white (negative brightness) to black (positive
        brightness). Contours are in units of 0.7~mJy per beam
        ($1\sigma$ level) from -9 to 0. Pixel size is 10~arcseconds. A
        smoothing by 3 pixels was applied. }
         \label{fig:rxj1}
   \end{figure}
   
   The SZ effect is clearly detected in one of the most X-ray luminous
   clusters, RXJ1347-1145, at a redshift of 0.45. The map shown in
   Fig.~\ref{fig:rxj1} is obtained after coadding 16 hours of rasters
   taken in January 1999. First results were described by
   Pointecouteau \etal (1999)~\cite{P99} and these observations are analysed at
   length by Pointecouteau \etal (2001)~\cite{P01}. A new mapping algorithm is
   used here in order to deal with the effect of wobbling (Marty \etal
   2001)~\cite{Marty}. A projected gas mass can be almost directly deduced from
   these observations $1.1\pm 0.1 \times 10^{14} M_\odot$ within an
   angular radius of $\theta = 74 ''$ in agreement with X-ray expected
   gas mass. The SZ effect is the strongest ever detected ($y=7\times
   10^{-4}$). This is accomplished with a high signal to noise (about
   20).

   Other clusters have been mapped with the same experiment and will
   be reported by Marty \etal (2001)~\cite{Marty}.

\subsection{Dark clusters}

\begin{figure}[htbp]
  \includegraphics[angle=90,height=.5\textheight]
    {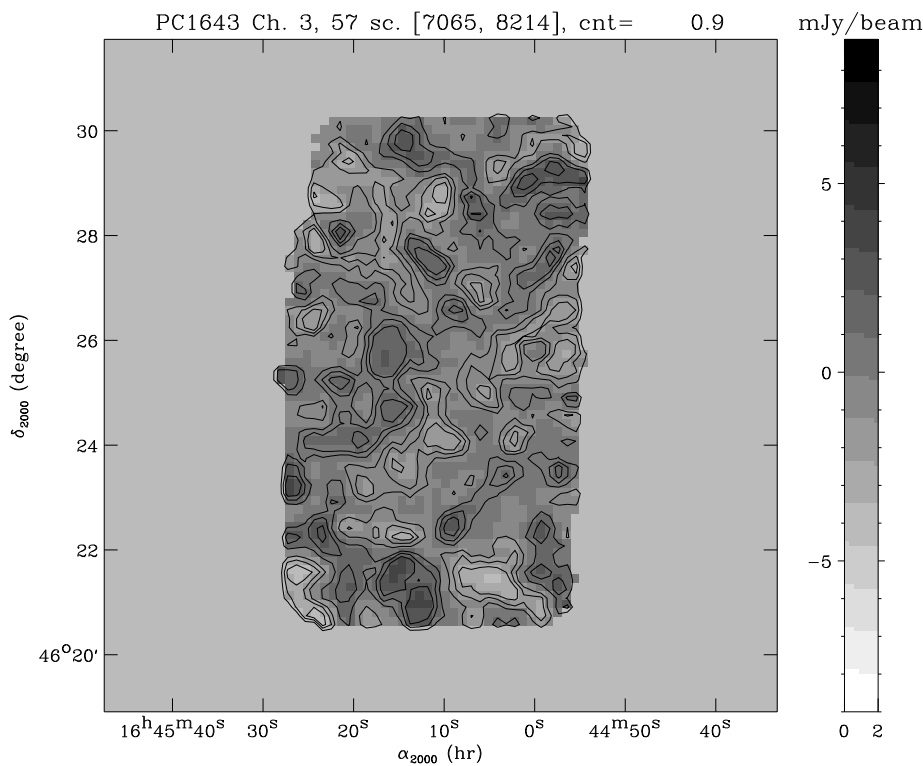}
      \caption{2.1~mm Diabolo map in the field of PC1643+4631. The grey scale is
        from white (negative brightness) to black (positive
        brightness). Contours are in units of 0.9~mJy per beam
        ($1\sigma$ level for the applied 30 arcsecond smoothing). The
        center of the Ryle decrement is at 16h45m11.2+46d24'56"}
         \label{fig:pc1}
   \end{figure}

\begin{figure}[htbp]
  \includegraphics[angle=0,height=.5\textheight]
    {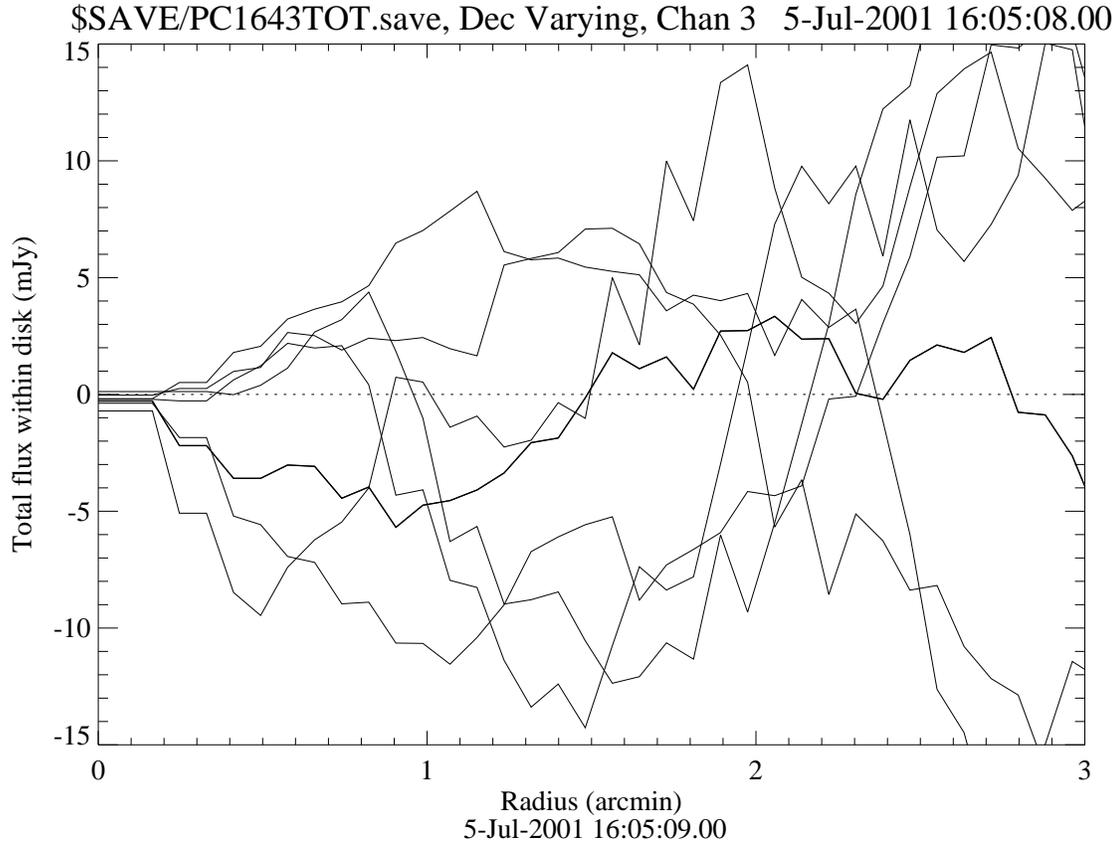}
      \caption{Integrated 2.1~mm flux radial profile obtained from the
        previous map. Each line corresponds to a central position with
        a constant right ascension but a declination varying by steps
        of 1 arcminute across the map.}
         \label{fig:pc2}
   \end{figure}
   
   A strong SZ effect has been detected with the radio Ryle
   interferometer at a position near the pair of quasars
   PC1643+4631 by Jones \etal (1997)~\cite{Jones}.  This brightness decrement observed at
   15~GHz could not be confirmed by other experiments like BIMA at
   28.5~GHz~\cite{Holzapfel} or SuZie at 2~mm~\cite{Ganga}. Here we
   wanted to map a sufficiently large map so as not to miss any
   decrement that could have been mispositioned, especially in
   declination, and have high resolution as well so as not to miss any
   relatively compact source.  About 18 hours were spent in January
   1999 providing our deepest field ever observed at 2.1~mm. 57 maps
   of 1200 seconds each can be coadded in order to have a typical
   sensitivity of 1.5~mJy ($1 \sigma$) for each of the 20 arcsecond
   pixel making up the final 4 by 9 arcmin map.  Fig.~\ref{fig:pc1}
   shows this final map. No strong SZ effect is detected in this map.
   To set a preliminary upper limit, we have computed the integrated
   flux inside varying radii for a given central position.
   Fig.~\ref{fig:pc2} shows that the absolute flux is never larger
   than 15~mJy in the 1 to 2 arcminute radius range (the optimum range
   for our 2.5 arcmin wobbling amplitude), whatever the center
   declination is chosen. We can safely exclude sources with an
   absolute flux larger than 20~mJy. The SZ effect expected with the
   minimum parameters advocated by Jones \etal~\cite{Jones} (core
   radius of 1~arcmin, $\Delta T/T =2\times 10^{-4}$) is -35~mJy at
   2.1~mm, which is not observed. If the Ryle decrement were due to a
   kinetic SZ effect, as would arise, for example, from a bubble of
   matter ionized by early quasars (Aghanim \etal~\cite{Aghanim}),
   then our spectral leverage implies a flux twice larger (-70~mJy)
   which is clearly not observed. Analysis of the PC1643 field in
   terms of CMB anisotropies should also be reported soon.

\subsection{2~mm source counts}
  From the previous deep survey, we also analysed the map for the
  presence of point sources. No sources could be detected at the level
  of 6~mJy in an area of 34~arcmin$^2$. A $2\sigma$ upper limit on
  integral number counts is thus $400\, \mathrm{deg^{-2}}$ at the flux
  limit of 6~mJy. This limit, which is above 850~\micron and 1.2~mm
  counts, could be improved with a larger observing time and/or next
  generation of bolometer instrument (see below). These constraints
  may prove very useful for the knowledge of high redshift galaxies.

\subsection{Sky noise}


\begin{figure}[htbp]
  \includegraphics[angle=90,height=.5\textheight]
    {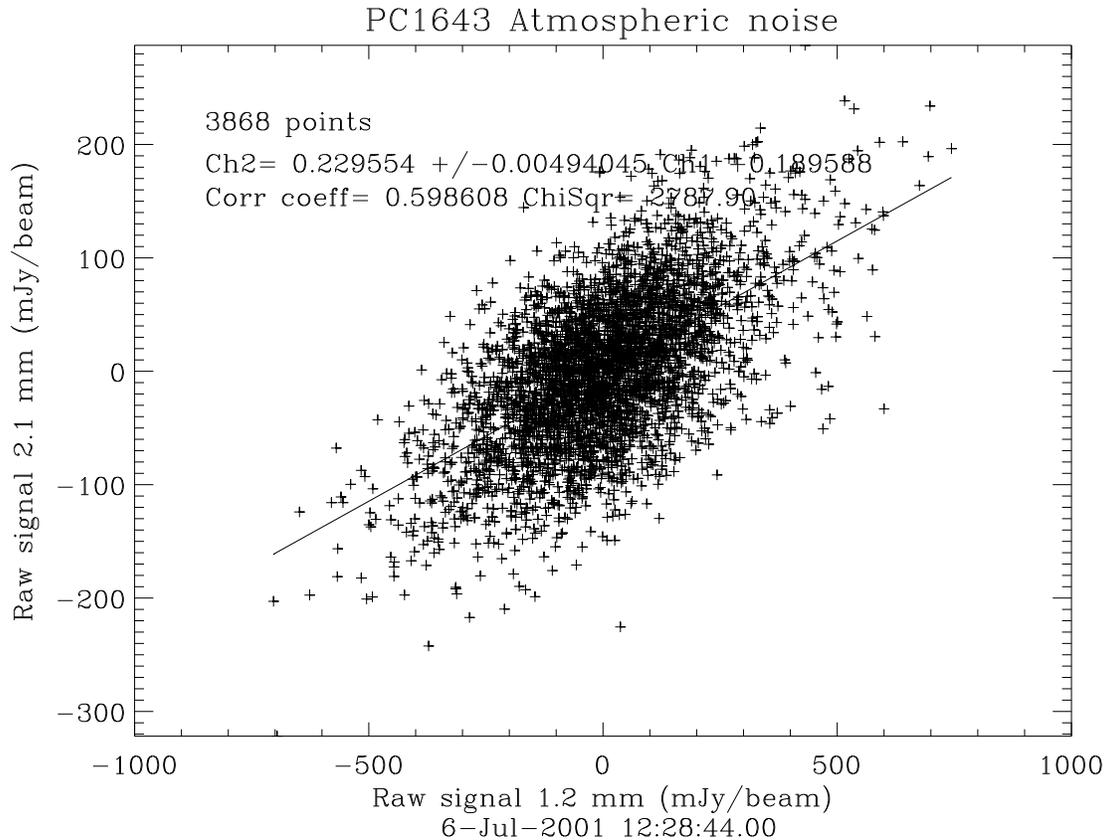}
      \caption{Illustration of millimetre sky noise.  
        Raw signal correlation between two bolometers during one scan
        made on the ``empty'' field PC1643. The signal at 1.2~mm
        (containing almost no SZ effect) provides a template on which
        the 2.1~mm (SZ) signal can be decorrelated. The slope is
        compatible with the spectrum expected from water vapour
        emission.}
         \label{fig:noise}
   \end{figure}
   
   One of the main limitations in ground-based (sub)millimeter
   observations is sky noise. This is due to inhomogeneous water
   vapour layers travelling above the telescope. This fluctuating
   emission produdes a spatially and temporally variable noise
   degrading the performance of millimetre continuum measurements. Two
   methods are used to counteract this noise. The spatial method is
   used when one wants to observe point sources with an array of
   bolometers, thereby one subtracts the average signal from
   neighbouring pixels (Kreysa, this conference). For extended sources
   this is insufficient. With a dual channel instrument like Diabolo
   we were able to perform the spectral method, whereby the SZ signal
   having a spectrum very different from the water vapour emission, a
   simultaneous measurement at 2 wavelengths (namely 1.2 and 2.1~mm)
   allows one to subtract the sky noise induced map at all spatial
   scales.  Fig~\ref{fig:noise} illustrates this method. In this case,
   the gain in signal to noise has proven not to be dramatic (30 to
   50\%). However the statistics of the signal is much improved in
   that the remaining (hopefully) detector noise is closer to
   Gaussian, hence improving the quality of SZ detections
   (\eg~\cite{Desert}).

\section{The step forward in millimeter continuum astrophysics}

The previous results have shown examples of the importance of the
large surveys at 1.2 but also 2.1~mm wavelengths with high resolution.
The 4 most important (extragalactic) reasons at 2.1~mm are the study
of the SZ effect in conjontion with X-ray observatory data, the study
of secondary anisotropies (the CMB is flat on these small angular
scales), the detection of primordial galaxies, and the mapping of
external galaxies. To achieve that goal, this conference has seen many
projects and realisations of cameras with many bolometers packed
together with individual horns. On the other hand, we wish to advocate
the use of filled arrays of bolometers fully sampling the available
focal plane of large millimetre dishes (see details
in~\cite{Camiram}). Indeed, there are three reasons for this new
design to be competitive: the sky noise may be better handled (no
instanteneous holes in the observed field of view), the confusion
noise, nearly reached even with large telescopes, can be better
tackled and the efficiency of photon gathering is optimised. The
challenge is clearly the optical behaviour of such new cameras (the
background is a million time larger than the objects to be detected),
adapting a multiwavelength operation (should we keep dichroic or use
wavelength sensitive piled up arrays?), multiplexing arrays of several
thousand pixels. Such cameras can be the workhorse for the
ground-based follow-up of large surveys made by the next generation
space instruments like SIRTF, Herschel and Planck. Developments in
France follow from the CEA/Leti design of a submillimeter camera for
Herschel (initially for SPIRE~\cite{Griffin} and now for PACS) and
other advance in NbSi technology~\cite{Dumoulin}. Prototypes are
currently being built and tested to qualify these new designs.





\begin{theacknowledgments}
  We thank INSU, MESR, and PNC for their continued support for the
  Diabolo experiment as well as IRAM for the observing logistics. The
  instrument could not have been made operational at the 30~m IRAM
  telescope without extensive testing on smaller millimeter
  telescopes, namely MITO (de Petris \etal, this conference) and POM2,
  a 2.5~m dish on Plateau de Bure operated by Bernard Fouilleux and
  Gilles Duvert (LAOG). We thank Marco de Petris and co-organisers for
  this very lively 2K1BC conference in front of the wonderful
  Cervinio.  Finally, we pay tribute to the memory of Guy Serra who
  was a pionneer in French and European submillimetre study of the
  diffuse galactic emission. Although he was very much involved in
  balloon experiments, he managed to help us in the alignment (by
  -15~deg.C) of the Diabolo instrument in the early days at Testa
  Grigia observatory (1994).
\end{theacknowledgments}

\end{document}